%% file: main.tex
\documentclass{article}

\usepackage{microtype}
\usepackage{graphicx}
\usepackage{subfigure}
\usepackage{booktabs} % for professional tables

\usepackage{hyperref}

\usepackage[accepted]{icml2025}

\usepackage{amsmath}
\usepackage{amssymb}
\usepackage{mathtools}
\usepackage{amsthm}

\usepackage[capitalize,noabbrev]{cleveref}

\theoremstyle{plain}

\theoremstyle{definition}

\theoremstyle{remark}

\usepackage[textsize=tiny]{todonotes}

\icmltitlerunning{Energy \& Force Regression on DFT Trajectories is Not Enough for Universal Machine Learning Interatomic Potentials}

\begin{document}

\twocolumn[
\icmltitle{Energy \& Force Regression on DFT Trajectories is Not Enough for Universal Machine Learning Interatomic Potentials}

% It is OKAY to include author information, even for blind
% submissions: the style file will automatically remove it for you
% unless you've provided the [accepted] option to the icml2025
% package.

% List of affiliations: The first argument should be a (short)
% identifier you will use later to specify author affiliations
% Academic affiliations should list Department, University, City, Region, Country
% Industry affiliations should list Company, City, Region, Country

% You can specify symbols, otherwise they are numbered in order.
% Ideally, you should not use this facility. Affiliations will be numbered
% in order of appearance and this is the preferred way.
% \icmlsetsymbol{equal}{*}

\begin{icmlauthorlist}
\icmlauthor{Santiago Miret}{il}
\icmlauthor{Kin Long Kelvin Lee}{il}
\icmlauthor{Carmelo Gonzales}{il}
\icmlauthor{Sajid Mannan}{iitd1}
\icmlauthor{N. M. Anoop Krishnan}{iitd1,iitd2}
% \icmlauthor{Firstname6 Lastname6}{sch,yyy,comp}
% \icmlauthor{Firstname7 Lastname7}{comp}
%\icmlauthor{}{sch}
% \icmlauthor{Firstname8 Lastname8}{sch}
% \icmlauthor{Firstname8 Lastname8}{yyy,comp}
%\icmlauthor{}{sch}
%\icmlauthor{}{sch}
\end{icmlauthorlist}

\icmlaffiliation{il}{Intel Labs, Santa Clara, California}
\icmlaffiliation{iitd1}{Department of Civil Engineering}
\icmlaffiliation{iitd2}{Yardi School of Artificial Intelligence, Indian Institute of Technology Delhi, New Delhi, India}

\icmlcorrespondingauthor{Santiago Miret}{santiago.miret@intel.com}
% \icmlcorrespondingauthor{Firstname2 Lastname2}{first2.last2@www.uk}

% You may provide any keywords that you
% find helpful for describing your paper; these are used to populate
% the "keywords" metadata in the PDF but will not be shown in the document
\icmlkeywords{Machine Learning Potentials, Graph Neural Networks, Geometric Deep Learning, Materials Science}

\vskip 0.3in
]

% this must go after the closing bracket ] following \twocolumn[ ...

% This command actually creates the footnote in the first column
% listing the affiliations and the copyright notice.
% The command takes one argument, which is text to display at the start of the footnote.
% The \icmlEqualContribution command is standard text for equal contribution.
% Remove it (just {}) if you do not need this facility.

\printAffiliationsAndNotice{}  % leave blank if no need to mention equal contribution
% \printAffiliationsAndNotice{\icmlEqualContribution} % otherwise use the standard text.

\begin{abstract}
Universal Machine Learning Interactomic Potentials (MLIPs) enable accelerated simulations for materials discovery. However, current research efforts fail to impactfully utilize MLIPs due to: 1. Overreliance on Density Functional Theory (DFT) for MLIP training data creation; 2. MLIPs' inability to reliably and accurately perform large-scale molecular dynamics (MD) simulations for diverse materials; 3. Limited understanding of MLIPs' underlying capabilities. To address these shortcomings, we aargue that MLIP research efforts should prioritize: 1. Employing more accurate simulation methods for large-scale MLIP training data creation (e.g. Coupled Cluster Theory) that cover a wide range of materials design spaces; 2. Creating MLIP metrology tools that leverage large-scale benchmarking,  visualization, and interpretability analyses to provide a deeper understanding of MLIPs' inner workings; 3. Developing computationally efficient MLIPs to execute MD simulations that accurately model a broad set of materials properties. Together, these interdisciplinary research directions can help further the real-world application of MLIPs to accurately model complex materials at device scale. 
 
\end{abstract}

\input{sections/1_intro}
\input{sections/2_background}

\input{sections/3_materials}
\input{sections/4_eval}

\input{sections/5_computation}

\input{sections/6_suggestions}

% % Acknowledgements should only appear in the accepted version.
% \section*{Acknowledgements}

% \textbf{Do not} include acknowledgements in the initial version of
% the paper submitted for blind review.

% If a paper is accepted, the final camera-ready version can (and
% usually should) include acknowledgements.  Such acknowledgements
% should be placed at the end of the section, in an unnumbered section
% that does not count towards the paper page limit. Typically, this will 
% include thanks to reviewers who gave useful comments, to colleagues 
% who contributed to the ideas, and to funding agencies and corporate 
% sponsors that provided financial support.

% \section*{Impact Statement}
% \vspace{-0.1cm}
% This paper presents work whose goal is to advance the field of  Machine Learning. There are many potential societal consequences of our work, none which we feel must be specifically highlighted here.

\bibliography{mlip}
\bibliographystyle{icml2025}

%%%%%%%%%%%%%%%%%%%%%%%%%%%%%%%%%%%%%%%%%%%%%%%%%%%%%%%%%%%%%%%%%%%%%%%%%%%%%%%
%%%%%%%%%%%%%%%%%%%%%%%%%%%%%%%%%%%%%%%%%%%%%%%%%%%%%%%%%%%%%%%%%%%%%%%%%%%%%%%
% APPENDIX
%%%%%%%%%%%%%%%%%%%%%%%%%%%%%%%%%%%%%%%%%%%%%%%%%%%%%%%%%%%%%%%%%%%%%%%%%%%%%%%
%%%%%%%%%%%%%%%%%%%%%%%%%%%%%%%%%%%%%%%%%%%%%%%%%%%%%%%%%%%%%%%%%%%%%%%%%%%%%%%
\newpage

\input{sections/appendix}
%%%%%%%%%%%%%%%%%%%%%%%%%%%%%%%%%%%%%%%%%%%%%%%%%%%%%%%%%%%%%%%%%%%%%%%%%%%%%%%
%%%%%%%%%%%%%%%%%%%%%%%%%%%%%%%%%%%%%%%%%%%%%%%%%%%%%%%%%%%%%%%%%%%%%%%%%%%%%%%

\end{document}

%% file: sections/1_intro.tex
\section{Introduction}

Machine learning (ML) opens up the possibility to greatly accelerate materials discovery for a vast range of applications, including the development of energy technologies to mitigate climate change, ubiqutuous computing technologies, and sustainable agriculture and manufacturing. 
The complexity of materials systems in modern applications continues to make empirical testing of devices, and their underlying materials, more and more challenging. As such, insilico materials simulation methods have become increasingly important to understand the properties and behavior of an growing set of diverse materials systems, ranging from solid-state crystals, molecules, and protein structures to name a few \citep{zeni2025generative, miret2024perspective, zitnick2020introduction, lee2023towards, terwilliger2024alphafold}. To fully realize the potential of insilico evaluation, it is necessary to consider how complex materials, which are comprised of large numbers of atoms governed by quantum mechanical laws, behave in diverse application conditions. Concretely, material systems are often part of multi-material devices that contain defects and imperfections while operating under various temperatures, pressures and other exogenous conditions. Therefore, similarly to how biological applications motivate the building of a virtual cell \citep{bunne2024build}, materials applications require the building of virtual devices (e.g. transistors, battery cells, nano-sized air filters) to accurately capture atomic interactions with quantum mechanical accuracy in complex environments.

\begin{figure*}[t]
    \centering
    \includegraphics[width=0.85\textwidth]{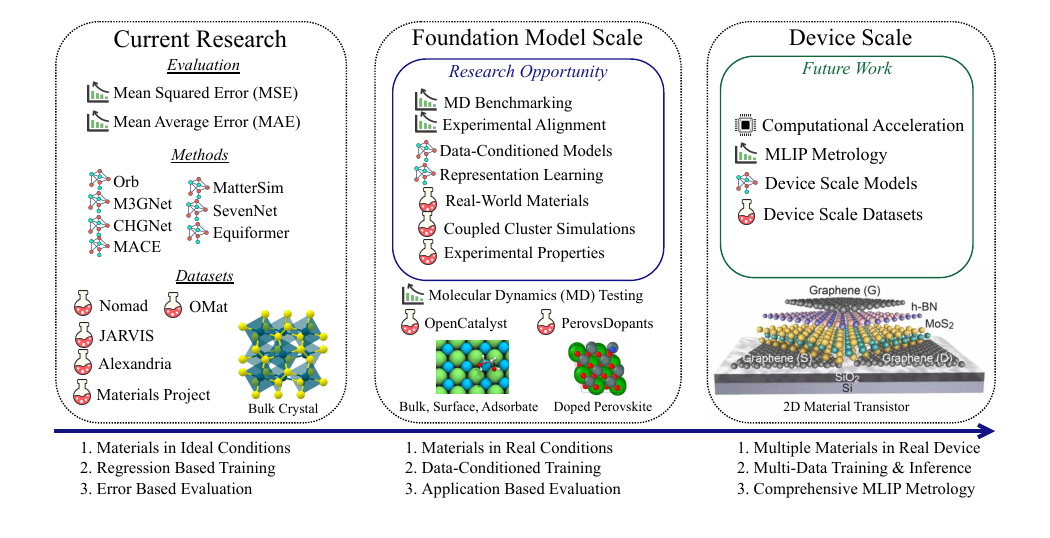}
    \vspace{-0.80cm}
    \caption{Overview of Machine Learning Interatomic Potentials (MLIP) requirements for device scale modeling. Current research focuses mainly on bulk structures in ideal conditions with regression-based training and error metric evaluation. To enable materials foundation models, we require higher quality training datasets that use more accurate simulation methods like Coupled Cluster Theory. MLIPs, in turn, should be evaluated as part of atomistic simulations of real-world materials in application-informed conditions (e.g., defects, standard temperature \& pressure, etc.). To reach MLIP-accelerated device scale modeling with quantum mechanical accuracy, we require new datasets and evaluation methods for complex materials systems in modern devices (e.g., 2D Material Transistors with multiple layers of distinct materials with specific functions), computational acceleration for proper inference, and comprehensive MLIP metrology.}
    \label{fig:mlip-req}
    \vspace{-0.50cm}
\end{figure*}

To date, the computational evaluation of materials systems with quantum accurate methods has mostly focused on applying approximations, such as Density Functional Theory (DFT), to idealized bulk structure materials \citep{jain2011high, yang2019predicting, saal2013materials, horton2019high, doerr2016htmd, miret2023the}. While these efforts have significantly improved the understanding of diverse materials, their applicability has been hindered by prohibitively high computational costs that scale exponentially with the number of atoms in the system. Modern ML methods for atomistic modeling, mainly based on geometric deep learning \citep{duval2023hitchhiker}, have emerged as a promising alternative to traditional computational chemistry pipelines. These ML methods boast the ability to model large material systems with quantum accuracy at constant compute cost following model training. This has led to the development of machine learning interatomic potentials (MLIPs), which are trained on large amounts of quantum simulation data. MLIPs aim to approximate the potential energy of diverse materials in a fast and accurate manner. In recent years, an increasing number of datasets \citep{chanussot2021open, tran2023open, barroso2024open, lee2023matsciml, schmidt2024improving, wang2024perovs, fuemmeler2024advancing} and models \citep{batatia2023foundation, neumann2024orb, yang2024mattersim, chen2022universal, deng2023chgnet} have been released, each claiming various degrees of generalizability for atomistic modeling of materials. While ongoing research has shown initial promise, significant work remains to realize the full potential of MLIPs in enabling physically accurate materials simulation at device scale.

We believe that for MLIPs to have real-world impact, current research priorities need to shift away from current benchmarks \citep{lee2023matsciml, riebesell2023matbench, choudhary2020joint, chanussot2021open} and place greater focus on enabling \emph{device scale} simulations.
Device scale simulation requires new algorithms beyond current training and inference methods that have mainly focused on predicting DFT trajectories for idealized bulk materials. Concretely, coupling large-scale MLIP inference with MD simulations can lead to faster simulations of larger, more complex materials in realistic application conditions. This in turn would allow practitioners to exctract more actionable insights from MLIP-driven simulations. To realize this interdisciplinary vision, we propose the following directions: 

\emph{1. Higher Accuracy MLIP Training Data Creation} 

As described in \Cref{sec:materials}, we require dedicated and collective efforts to create datasets and benchmarks that better represent the complexity of real-world materials applications. These new datasets should be both chemically and structurally diverse, and cover a wide range of physical conditions (e.g. temperatures, pressures, defects) and phase changes (e.g. solid-to-liquid, liquid-to-gas). 
Data generation methods should be reproducible and verifiable for both simulation and experimentally measured data. As we discuss in \Cref{sec:materials}, nuances with popular DFT approximations, methodologies, and implementations lend themselves to large variances in simulation results, thereby making MLIPs trained solely on DFT data less reliable. To improve training data quality, we recommend using higher accuracy simulation methods like Coupled Cluster Theory.

\emph{2. MLIP Metrology -- Interpretable and Materials Science Informed MLIP Evaluation}

As described in \Cref{sec:eval}, MLIP evaluation methods should capture relevant behavior for MLIPs applied to MD simulations that model experimentally measured properties. We also require MLIP metrology methods that enable better interpretability of MLIPs to highlight their capabilities and limitations. Taken together, these efforts will enable greater understanding of MLIP methods and utilities, while also fostering trust in the materials science community.

\emph{3. Computationally Efficient MLIP Inference Workflows for Device-Scale Materials Simulation}

As described in \Cref{sec:computation}, MLIPs require efficient inference when deployed in MD simulations to disrupt traditional simulation methods and unlock the promise of device scale modeling. Faster inference will enable MLIPs to model larger and more chemically complex systems compared to classical methods. However, this will require significant engineering advances, including: re-usable computation kernels, optimized architectures, and scalable inference methods deployed in simulation software stacks.

%% file: sections/2_background.tex
\section{Background \& Current Work} \label{sec:background}

\begin{figure}[h]
    \vspace{-0.2cm}
    \centering
    \includegraphics[width=\linewidth]{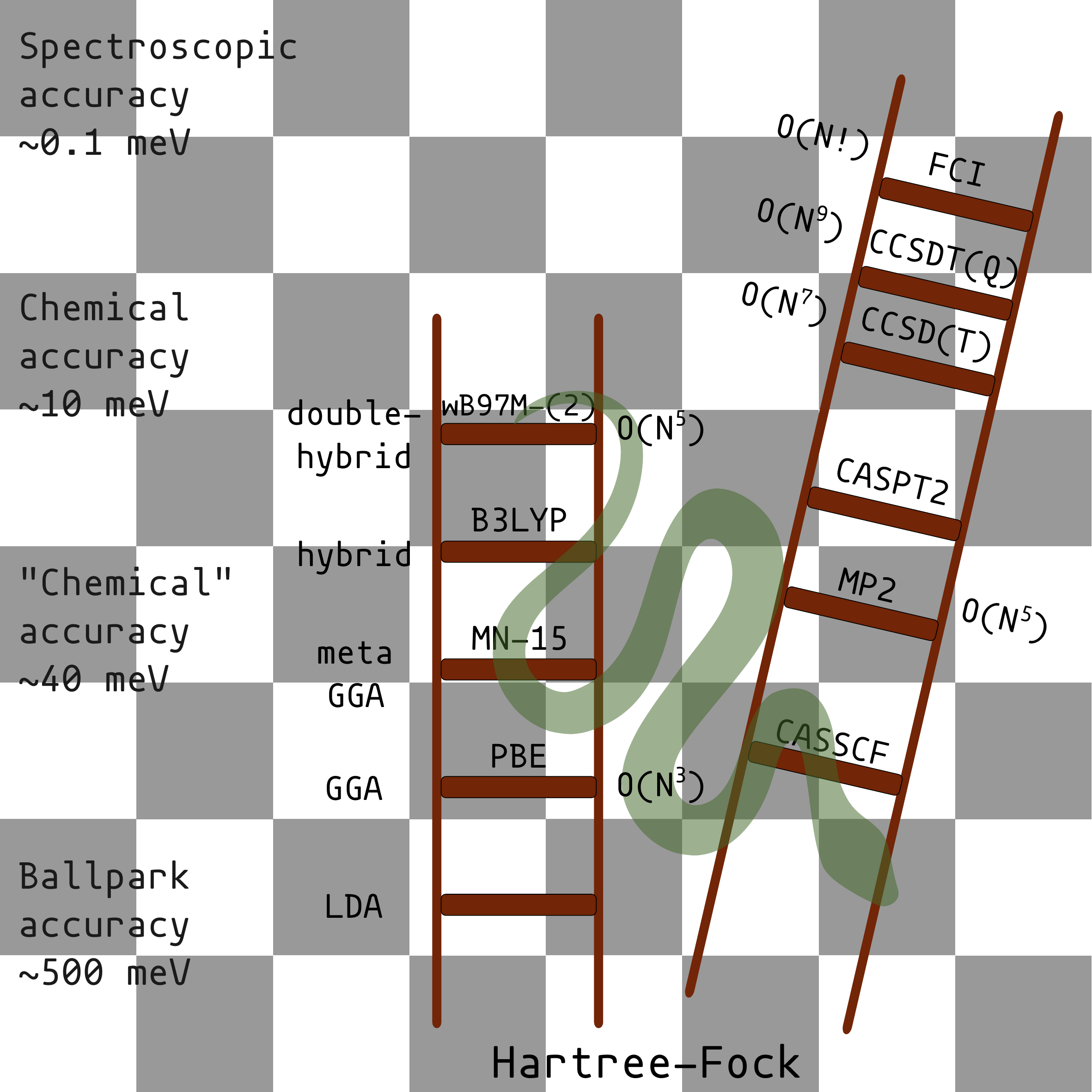}
    \vspace{-0.65cm}
    \caption{An artistic interpretation of Jacob's ladder \citep{perdew2005prescription}, extended to include wavefunction methods. Going up the ladders improves both accuracy and precision, but is generally proportional to increased time complexity. We stress that, while there are many nuances associated with method choice for atomistic systems, particularly with multireference methods like CASSCF \citep{roos1980complete}, these accuracy trends are generally observed.}
    \label{fig:jacobsladder}
    \vspace{-0.2cm}
\end{figure}

Most atomistic ML methods are based on geometric deep learning architectures \citep{duval2023hitchhiker}, which provide a useful framework for encoding inductive biases, such as invariance, equivariance, and other symmetries. Here, we provide an overview of relevant quantum mechanical simulation methods for training data generation, which are graphically depicted in \Cref{fig:jacobsladder} with Jacob's ladder of modeling accuracy. \Cref{app:aimd} details how MLIP inference is coupled with MD simulation for materials modeling.

\textbf{Density Functional Theory (DFT):} DFT  is a widely used quantum mechanical framework to model the behavior of electrons in atomistic systems. Conventionally, DFT aims to solve the non-relativistic electronic Schr\"{o}dinger equation by modeling the electron density from which all other electronic properties can be derived (e.g. potential energy, multipole moments). A hypothetical functional $F$, of the electron density, is estimated by iteratively minimizing its corresponding electronic energy, thereby producing a variational solution. \citet{jonesDensityFunctionalTheory2015} provides a comprehensive review of DFT fundamentals, which primarily rely on the assumption that an exact functional can perfectly describe the electron density for all and any point cloud of atoms in 3D space.  This exact functional is unknown---some regard it as \emph{unknowable} \citep{schuch2009computational}---and in practice is approximated using an exchange-correlation term \citep{kohn1965self}. \footnote{The theoretically exact framework for DFT only holds true for a very small number of cases \citep{mayerConceptualProblemCalculating2017}. In alignment with the literature, we use DFT to refer to the approximation.} Two of the most commonly used functionals include the Perdew-Burke-Ernzerhof functional (PBE) \citep{perdew1996generalized} and Becke, 3-parameter, Lee-Yang-Parr functional (B3LYP) \citep{leeDevelopmentColleSalvettiCorrelationenergy1988,beckeDensityfunctionalExchangeenergyApproximation1988}. These functionals have been applied in many popular ML materials datasets, such as Materials Project \citep{jain2013commentary}, Alexandria \citep{schmidt2024improving}, and OpenCatalyst \citep{chanussot2021open}. Different approximations have subsequently led to families of density functionals for diverse cases. As such, the \emph{choice} of approximations and parameterizations ultimately affects the performance of the functional on specific elements of the periodic table and/or classes of molecules and materials \citep{apra2020nwchem, kuhne2020cp2k, giannozzi2009quantum}. While DFT, which scales $\mathcal{O}(N^3-N^5)$ with $N$ electrons, has been popular for training current MLIPs, it has known inaccuracies, inconsistent results, and other shortcomings as described in \Cref{sec:materials}.

\textbf{Coupled Cluster (CC) Theory:} CC methods \citep{cizekCorrelationProblemAtomic1966} employ highly accurate ``wavefunction'' or \textit{ab initio} quantum mechanical formulations that yield systematically improvable results. CCSD(T) \citep{purvisFullCoupledclusterSingles1982}, which applies single, double, and perturbative triple excitations is regarded as the ``gold standard'' of quantum chemical methods. The hallmarks of CC-based methods is the ability to truncate a Taylor series of excitation operators to systematically choose between computational cost and accuracy---this series converges to the exact solution to the non-relativistic Schr\"{o}dinger equation (\Cref{fig:jacobsladder}) for fixed nuclei, as opposed to DFT methods that involve semi-empirical tuning. Notably, MLIP evaluation today is often done with DFT references while new DFT functionals are benchmarked against CCSD(T), which in turn is usually compared against experimental observables. The computational cost of CCSD(T), which scales $\mathcal{O}(N^7)$ with $N$ electrons, has limited its application to small molecules until very recently \citep{tang2024approaching}, where hybrid methods with MLIPs have enabled large scale, finite-temperature simulations of periodic systems with CC quality \citep{herzogCoupledClusterFinite2024}.

%% file: sections/3_materials.tex
\section{MLIPs for Materials Science Simulations} \label{sec:materials}

%%%%%%%%%%%%%%%%%%%%%%%%%%%%%%%%%%%%%%%%%%%%%%%%%%%%%%

Current gaps in MLIP research include the reliance on DFT for training data generation (\Cref{sec:dft-limits}) and the underrepresentation of diverse materials in MLIP datasets and evaluation methods (\Cref{sec:mat-limits}).
\Cref{sec:mat-limits} also describes the modeling of materials under realistic conditions, including how applying MLIPs jointly with MD simulations enables benchmarking against experimentally measured properties. \Cref{sec:mlip-dev} outlines recommendations for new research directions towards future MLIP training and development. 

\subsection{Limitations of DFT for Training Data Generation} \label{sec:dft-limits}
\textbf{DFT Methods Have Limited Accuracy:} A critical aspect for dataset quality is the quality of the method used to obtain ground truth labels. While DFT methods are by far the most common source of data for both ML research and materials scientists alike, the implementations used rely on a hierarchy of approximations. The accuracy of DFT rests primary on the quality of the functional form (see \Cref{sec:background}), and its shortcomings in the description of many \emph{key} chemical and physical phenomena are well-documented throughout the literature \citep{schuch2009computational}. The weaknesses of DFT manifest themselves in multiple forms, such as inaccuracies in calculating band gaps \citep{perdewDensityFunctionalTheory1985,bystromAddressingBandGap2024a}, fractional charges \citep{cohenChallengesDensityFunctional2012}, and general systems that demonstate static/strong electron correlation \citep{cohenFractionalSpinsStatic2008,suDescribingStrongCorrelation2018}. Recent DFT benchmarking work by \citet{araujoAdsorptionEnergiesTransition2022} shows that, without intricate corrections, the commonly applied PBE+D3 functional results in errors on the order of tens of kcal/mol (${\sim}0.5$\,eV) for adsorption energies on transition metal surfaces, making it impossible to model catalytic activity with uniform accuracy across the periodic table. Given the parametric approximations required for DFT, it is also easy to significantly overfit when modeling energy \citep{medvedevDensityFunctionalTheory2017}, resulting in poor generalization to other material properties. These errors can then further propagate to MLIP-based property prediction models.

\textbf{DFT Introduces Inconsistencies and Reproducibility Issues Across Different Codes:} 
The reproducibility of DFT calculations across different electronic structure codes presents a significant challenge for MLIP data generation. Even when using identical exchange-correlation functionals, variations in implementation details, basis sets, pseudopotentials, and numerical parameters can lead to discrepancies in computed energies and forces \citep{schuch2009computational,bootsmaPopularIntegrationGrids2019}. These inconsistencies, as prominently displayed in \citet{lejaeghere2016reproducibility,bosoniHowVerifyPrecision2024}, become particularly problematic when training MLIPs, as the resulting models inherently encode code-specific biases---not necessarily physical behavior. Another challenge is the DFT data generated by open-source code \textit{viz-a-viz} closed source packages. Most public DFT datasets (and workflows to generate those), such as MPTrj, OMat24 \citep{barroso2024open}, and Alexandria, rely on Vienna Ab-initio Simulation Package (VASP) \citep{kresse1994ab}, a closed-source commercial package. We encourage the use of open-source codes, such as Quantum Espresso \citep{giannozzi2009quantum} and CP2K \citep{kuhne2020cp2k}, to make data generation more accessible and reproducible. Even though these methods still suffer from implementation-specific variance, they enable greater scientific transparency.

\textbf{Data Generation with Higher Accuracy Methods:} Overall, we argue that the utility of continuing to apply DFT for large-scale data generation has diminishing returns due to known limitations and inaccuracies of the method. As such, we advocate for changing data generation methods to higher accuracy methods to avoid running into previously encountered obstacles with ML for atomistic modeling. Applying DFT for data generation and MLIP training may still be useful in targeted cases that further the understanding of specialized systems \citep{wang2024perovs, louIntelligibleModelsClassification2012} or novel scientific vantage points for materials science. However, MLIP performance is ultimately gated by the availability of high quality data, which necessitates more accurate simulation methods and targeted real-world experiments. Otherwise, many of the machine learning discoveries may be prone to hacking \citep{ghugare2024searching, govindarajan2024crystal} or materials with limited experimental utility \citep{cheetham2024artificial}. In our opinion, perhaps the strongest advantage of MLIPs is the ability to improve physical accuracy at constant computational inference cost: if an MLIP architecture approximates DFT [$\mathcal{O}(N^3-N^5)$], the true benefit lies in approximating a higher quality, higher cost function with the same number of floating point operations such as CCSD(T) [$\mathcal{O}(N^7)$], or even full configuration interaction [$\mathcal{O}(N!)$] as exact solutions to the non-relativistic electronic Schr\"{o}dinger equation. Given the computational cost of generating these labels, more research is required to effectively bootstrap high volume, low quality data (DFT) into low volume, high quality data (CCSD(T)). 

\textbf{Challenge 3.1.1.} \textit{Develop targeted higher accuracy simulation datasets (e.g., CCSD(T)) to enable the training and evaluation of MLIPs on higher quality data. The dataset generation methods should employ open-source ab-initio packages to enhance transparency, repoducibility, and accessibility to avoid previsouly encountered challenges with large-scale dataset generation based on DFT.}

\textbf{Hybrid ML+QM Approaches:} One interesting related research direction consists of ``hybrid'' solutions that aim to improve how quantum mechanical models are solved with machine learning: parameterized models are used as a basis or ans\"{a}tze for the solution of classical methods; i.e. bounded solutions to the exact Schr\"{o}dinger equation. A strong example of this approach is the variational Monte Carlo approach from \citet{pfauAccurateComputationQuantum2024} built on top of Psiformer/FermiNet \citep{pfau2020ferminet,glehnSelfAttentionAnsatzAbinitio2023}. These methods show significant promise for yielding accurate models for electronic and nuclear properties beyond simple energy and force regression targets. The difficulty faced by these approaches mainly lies in domain adaptation (e.g. periodic boundary conditions for solid-state structures) and scaling (i.e. larger and more diverse atomic systems). An important distinction to emphasize is that one quality target noted in \citet{pfauAccurateComputationQuantum2024} is based on \emph{experimental} measurements. 
When benchmarking MLIPs, experimental observations should be one of the most important criteria, even though these observations contain pertinent uncertainties. For example, spectroscopic and spectrometric methods remain the only way to infer the success of materials synthesis using incomplete signals based on pattern matching. This means that if a material has certain spectroscopic signatures, then it has a high chance of being the material of interest. 
As research in MLIPs progresses, the application of ML models for materials science should aim to reproduce or \emph{predict} the same expected signatures \citep{cheng2024determining}.

\textbf{Challenge 3.1.2.} \textit{Develop hybrid ML+QM methods that improve predictive accuracy against experimental data, such as observable properties and spectroscopic signatures.}

\subsection{Exploring Broader Ranges of Materials Under Realistic Application Conditions} \label{sec:mat-limits}

\begin{figure}[h]
    \vspace{-0.35cm}
    \centering
    \includegraphics[width=0.98\columnwidth]{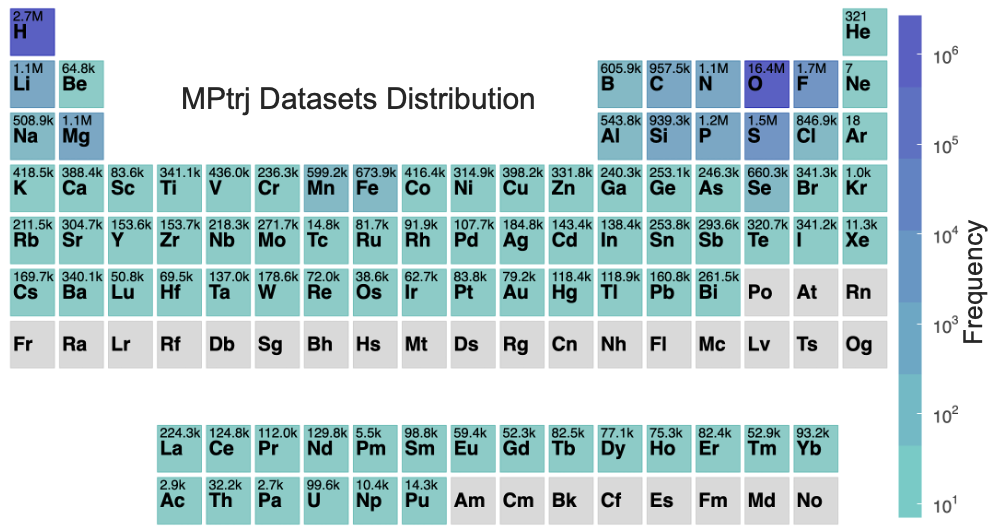}
    \vspace{-0.35cm}
    \caption{Frequency of elements in MPtrj dataset. The color bar in the figure represents a logarithmic scale ranging from low to high values, with the corresponding numbers indicating the frequency of each element's presence in the MPtraj dataset.}
    \label{fig:mptrj-data}
    \vspace{-0.35cm}
\end{figure}

Current large datasets, such as MPtrj \citep{deng2023chgnet}, which several universal potentials have been trained on, have primarily focused on a limited set of materials. They sometimes even neglect broad classes of materials present in real-world applications, such as metallic glasses, disordered materials, metal organic frameworks, polymers, alloys, and doped semiconductors \citep{burner2023arc, wang2024perovs, vita2023colabfit, downs2003american}. Moreover, the current datasets like MPtrj are biased towards specific families of materials and elements. \Cref{fig:mptrj-data} shows that certain elements such as H, Li, Mg, Si, P, and O, along with their possible compounds, are overrepresented, with 89 elements completely missing. This highlights a significant gap in data availability for many materials systems, many of which are relevant to real-world applications. 

In addition to the aforementioned limitations on DFT accuracy, the limited set of systems studied imposes additional constraints on MLIP models generalizing to new designs. This is further compounded by many DFT calculations only being evaluated in ideal conditions, meaning zero temperature and pressure, which does not properly approximate most application conditions. Some initial work has shown promise in ML models generalizing across different temperatures, pressures, and excited states \citep{merchant2023scaling, batatia2023foundation, westermayr2020machine}, but further work remains in understanding the abilities of MLIPs to model materials across a diversity of relevant application conditions. Variations may include temperature, pressure, inclusion of defects, and phase changes to name a few.

\textbf{Interface and Multi-Material Interactions:}
Most DFT datasets and benchmarks to date, with a notable exception of OpenCatalyst \citep{chanussot2021open, tran2023open}, have focused on modeling the properties of structures that represent a single bulk crystal or small molecules. While valuable to bootstrap the development of MLIPs, this is not sufficient to enable MLIP-based device scale simulation. Materials in modern devices interact with other materials around them to fulfill various complex performance requirements. As shown in \Cref{fig:mlip-req}, a modern transistor requires the intricate combination of multiple materials \citep{reddy2022comprehensive}, each of which provide essential functions. The need to model multiple materials and their interfaces reliably and accurately introduces a significantly more complex task than what is available in current training datasets.

\textbf{Aligning MLIP-Based Simulation Evaluation Towards Experimental Properties:} Given the vast set of applications for different materials, it is important to be able to model diverse sets of properties. One of the main advantages of MLIP-based simulations is the ability to model experimentally measurable properties, such as elastic moduli, thermal expansion, and thermal conductivity. Moreover, while considering the experimental properties, aligning the corresponding synthesis and testing conditions are important to make a meaningful comparison with the corresponding simulations. Thus, the measurements should be aligned to real-world conditions and properly documented. Such a database could potentially serve two purposes: i) Benchmarking MLIPs based on realistic scenarios encountered during applications; ii) Applying experimental data to train MLIPs for more accurate property prediction. Differentiable simulation frameworks provide an interesting framework to further the development of MLIPs by back-propagating directly through simulation trajectories to update MLIP and simulation parameters. Given the nascense of differentiable simulation frameworks, more research is needed to develop performant and rigorous simulation and theoretical frameworks \citep{gangan2024force, metz2021gradients}.  

\textbf{Challenge 3.2.} \textit{Systematically map and address data sparsity in materials databases through strategic generation of datasets for underrepresented chemical spaces and material families, enabling more comprehensive coverage of the materials genome in real-world conditions. Scale data generation to complex materials systems involving interfaces, reactions, and other complex interactions while driving greater aligment to experimental measurements.}

\subsection{MLIP Training \& Representation Learning} \label{sec:mlip-dev}

Most of today's MLIPs are based on message passing graph neural networks (GNNs) with geometric inductive biases that infuse different types of symmetries \citep{duval2023hitchhiker}. Many of these models have been trained and evaluated using regression objectives for energy and forces or other materials properties \citep{riebesell2023matbench, chanussot2021open, choudhary2020joint, lee2023matsciml}. Recently, new models have emerged that have made use of multiple datasets in their training pipeline \citep{barroso2024open, neumann2024orb} to achieve better overall performance and generalization. Compared to other fields, few methods have been proposed for self-supervised MLIP pretraining \citep{DeNS}. The success of denoising-based pretraining for effective representation learning in adjacent fields, such as proteins \citep{abramson2024accurate, zhang2023protein} and small molecules \citep{zaidi2023pretraining}, as well as the increasing data diversity related to MLIPs create the need to explore effective representation learning methods. 

As the scale of datasets increases, the computational requirements of pretraining will also increase, thereby prompting further investigation into scaling laws for MLIPs \citep{frey2023neural}. As both training and inference scale requirements increase, the utility of inductive biases will continue to remain a pertinent question. Recent work indicates that GNNs provide useful inductive biases \citep{alampara2024mattext} even though conflicting evidence has emerged in adjacent fields. Further, in the limit of large data regimes, invariant GNNs may perform as well as equivariant ones at lower cost~\cite{qu2024importance,brehmer2024does, JMLR:v25:23-0680}. Thus, a critical analysis of the inductive biases required while considering scalability and accuracy is needed to identify optimal architectures. Message passing in GNNs, for example, can become a bottleneck when deployed in simulations given the iterative nature of the inference pass needed to cover large graphs. This has prompted architectures without message passing that provide similar performance as GNNs \citep{bochkarev2024graph}. Additionally, given the tight coupling between MLIPs and data generation methods, the field could also benefit from data-centric ML research related to designing datasets and continued model improvement \citep{oala2024dmlr}, as well as knowledge distillation described in \Cref{app:distillation}.

\textbf{Challenge 3.3.} \textit{Creation of MLIPs architectures with relevant feature engineering, inductive biases, and training methods. Novel MLIPs should utilize larger and more heterogenous datasets, potentially drawing on ideas from large-scale representation learning for greater generalization.} 

%% file: sections/4_eval.tex
\vspace{-0.2cm}
\section{MLIP Metrology: Testing \& Interpretability} \label{sec:eval}
\vspace{-0.1cm}

%%%%%%%%%%%%%%%%%%%%%%%%%%%%%%%%%%%%%%%%%%%%%%%%%%%

Since MLIPs serve as materials design tools for scientific end-users, they can benefit from being intuitive, reliable, and easy to test and analyze. As such, the development of \emph{MLIP metrology} to reliably analyze properties, behavior, and limitations for MLIPs becomes important. Based on current research, we suggest a preliminary start of MLIP metrology techniques, which help build towards a greater understanding of the capabilities and limitations of MLIPs, focused on: 1. large-scale benchmarking across materials and conditions as described in detail in \Cref{app:sec:mlip-eval}; 2. MLIP visualization and analysis, such as energy landscape visualization \citep{bihani2024lowdimensional}; 3. stability based analysis methods to enhance simulation reliability \citep{brandstetter2022message,raja2024stability,ibayashi2023allegro}; 4. interpretability studies to understand MLIP inner workings \citep{leeDeconstructingEquivariantRepresentations2024}. 

\textbf{Challenge 4.1} \textit{Develop MLIP metrology relying on large-scale evaluation of MLIPs in diverse materials, visualizing the energy landscape instead of simple pairwise interactions, and analyses of stability during MD simulations.} 

In the case of interpretability, for example, recent work has shown a lack of understanding and transparency in the learned representations of equivariant models using spherical harmonics and tensor products commonly used in current MLIPs. \citet{leeDeconstructingEquivariantRepresentations2024} showed that latent embeddings of an equivariant model projected and visualized with manifold structure preserving methods like PHATE \citep{moonVisualizingStructureTransitions2019} showed no discernable structure, highlighting a strong need for both new projection methods \emph{and} control over training dynamics for these methods. One potential architectural remedy relates to ``white-box'' or ``glass-box'' models, named so due to their intrinsic transparency and comprehensibility (i.e. intentionally simple structure) or through post-hoc explanation \citep{esdersAnalyzingAtomicInteractions2025,goethalsNonlinearNatureCost2022, esders2024analyzing, wangX2GNNPhysicalMessage2024}. While there have been traditional modelling efforts dedicated to comprehensible models, they receive significantly less attention to their black-box counterparts. \citet{pfauAccurateComputationQuantum2024} provides a framework for showing state-of-the-art modelling whilst remaining highly intuitive to computational chemists by infusing chemical first principles into the neural network. 

An important consequence of the paradigm proposed by \citet{pfauAccurateComputationQuantum2024} is the ability for learned representations and intermediate solutions to pass property tests, in contrast to black-box MLIPs: eigenvalues and vectors from derived solutions behave as physical models do (e.g. electron spin), and in the context of comprensiblilty, the ability to obtain \emph{variational} bounds on results means that quantitative behavior is well-understood. Intermediate approaches towards comprehensible models for materials modeling could borrow ideas from interpretable modeling choices, such as generalized additive models (GAM) \citep{hastieGeneralizedAdditiveModels1986,woodGeneralizedAdditiveModels2024} and derivatives like explainable boosting machines \citep{louIntelligibleModelsClassification2012}. 
An example hybrid approach that marries existing approaches with interpretable methods could be mapping equivariant features embedded as irreducible representations to outputs with GAMs---in doing so, we preserve physically motivated and intuitive signals (i.e. learned features with specific symmetries), and are able to understand \emph{why} a prediction is made through decomposition. Relating to \citet{leeDeconstructingEquivariantRepresentations2024}, this may unlock further improvements in model design and hyperparameter choice.

\textbf{Challenge 4.2} \textit{Develop interpretable architectures and methods that probe MLIPs, making them transparent and interpretable, either by design or post-hoc analysis.}

%% file: sections/5_computation.tex
\vspace{-0.2cm}
\section{Computational \& Modeling Considerations} \label{sec:computation}
\vspace{-0.1cm}

While tools exist today to integrate MLIPs into materials simulation codes \citep{gupta2024kusp}, such as ASE \citep{HjorthLarsen_2017} and LAMMPS \citep{thompson2022lammps}, the evaluation of MLIPs on simulation benchmarks remains limited in large part due to inefficient computation \citep{gonzales2024benchmarking}. Some architectures, such as MACE \citep{batatia2023foundation}, have achieved additional acceleration by implementing the model architecture using the Kokkos language \citep{9485033}. While re-implemtation might be feasible in isolated cases, this approach may not be scalable given the dominance of PyTorch for MLIP development and training.  
One promising approach for scaling the training and deployment of MLIPs is the emergence of performant, cross-hardware programming language like JAX \citep{jax2018github} and Triton \citep{tillet2019triton} that encourage community driven development. These langauges have already enabled the development of MLIP-based simulation frameworks \citep{schoenholz2020jax, Hu2020DiffTaichi:, helal2024mess, doerr2021torchmd} along with faster training and inference of MLIP architectures based on scalable primitives \citep{lee2024scaling}.

\begin{figure}[ht]
    \centering
    \includegraphics[width=\linewidth]{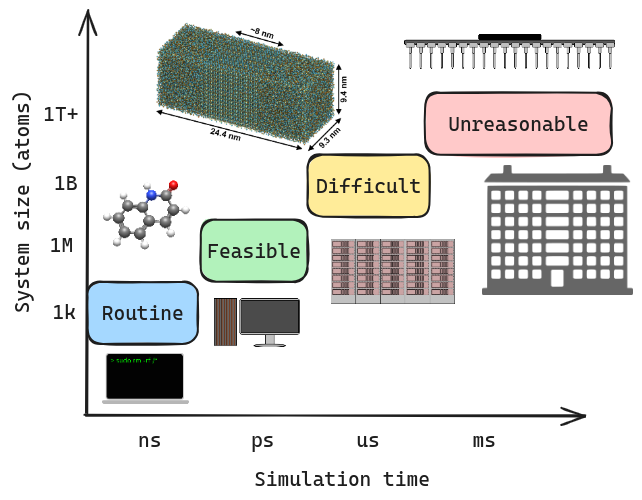}
    \vspace{-0.90cm}
    \caption{Schematic of length and time scales relevant to materials modeling. Annotations indicate approximate computational requirements; each block region corresponds to the amount of effort required for the corresponding scale of compute within a ``timely'' fashion. Routine and feasible refer to on the order of hours to days, while difficult can extend from weeks to months depending on the scale of distributed computing (e.g., the 2023 Gordon Bell prize submission by \citet{kozinsky2023scaling}). ``Unreasonable'' means technically possible, but practically improbable due to to collective compute and engineering efforts required, and corresponds to the device scale of an Intel 8086--a microprocessor from the late 1970's--assuming a representative ${\sim}$30,000 atoms per transistor.}
    \label{fig:compute-scales}
    \vspace{-0.60cm}
\end{figure}

As shown in \Cref{fig:compute-scales}, hardware and software advancements have reached new levels of maturity, thereby enabling the deployment of higher accuracy methods described in \Cref{sec:materials}. While these advancements show promise for better data generation methods, a significant gap remains for effective MLIP deployment. Like many other computational fields, there are physical effects and interactions that can only be modeled over large spatial and temporal scales---in the case of materials science, logarithmic in the number of atoms \emph{and} the time range. To date, pragmatic modeling choices are made to enable computationally \emph{tractable} but not necessarily realistic simulations to be conducted. 

The majority of the field currently focuses on pure crystalline structures that take full advantage of small unit cells, but fall short of the device-scale regime that requires the explicit treatment of billions of atoms composed into functional components (e.g. transistors), with timescales spanning from femtoseconds (e.g. electric field response) to microseconds or more (e.g. thermal dissipation, diffusion driven reactive chemistry, long-lived excited states). While MLIPs undoubtedly benefit from better time complexity scaling than conventional methods, \Cref{fig:compute-scales} shows that the forefront of atom scaling is still barely within reach of historical computing devices like the Intel 8086 with tens of thousands of transistors composed of hundreds of millions of atoms \citep{kozinsky2023scaling}, let alone modern devices with hundreds of billions. Using the empirical strong scaling results on $10^8$ atoms by \citet{kozinsky2023scaling} with 1024 nodes of Nvidia A100, reaching one millisecond of simulation time would require ${\sim}$105 years of dedicated computation. Thus, achieving the required scale needed to simulate properties of interest in integrated devices would require significant algorithmic and/or hardware improvements: performant kernels hundreds if not thousands of times faster could render at least the temporal aspect ``difficult'', although the atomic scale would remain in question. Hardware and algorithm co-design has enabled breakthroughs before, such as Anton for MD simulations \citep{shaw2021anton}. 

Another interesting direction is on fundamental changes to how MD simulations are performed. Notably, recent efforts such as TimeWarp \cite{klein2024timewarp} and distributional graphnormer \cite{zheng2024predicting}, have adapted generative modeling methods to improve sampling of rare and notable dynamics and minimize nominally wasteful timesteps. We note, however, that these approaches are still in their infancy and have been developed primarily for small molecules and structural biology, and thus have yet to be proved at the scale of hundreds of millions of atoms. Nonetheless, they appear to be highly promising research avenues.

\textbf{Challenge 5.1} \textit{Scalable methods for MLIP deployment on large-scale atomistic systems leveraging tightly coupled hardware/software co-design, and performant differentiable software stacks that drive new MLIP-enabled capabilities.}

%% file: sections/6_suggestions.tex
\vspace{-0.2cm}
\section{Alternative Viewpoints \& Approaches}
\vspace{-0.1cm}
\textbf{AV1: DFT is Good Enough \& MLIPs Are Already Universal.} One reasonable alternative viewpoint is that current methods that aim to replicate DFT-level accuracy for property prediction provide enough functional accuracy to perform large-scale materials screening leading to a downselection mechanism for further analysis. From that perspective, the main goal of MLIPs is to provide an accelerated way of filtering material candidates through reasonable property prediction and reduce the number of required downstream experiments. Additionally, MLIPs and DFT validation have already shown their effectiveness in validating the predicting of diverse materials generative models \citep{merchant2023scaling, levy2024symmcd, jiao2024space, zeni2025generative, miller2024flowmm, ding2024matexpert, gruver2024finetuned} that could provide valuable input to materials designers. As described in \Cref{sec:dft-limits}, DFT generally provides a reasonable trade-off between accuracy and computational cost, which largely underlies its popularity for high-throughput simulation and data generation. On top of that, there are cases where DFT is known to simulate properties well, informing both materials discovery and understanding. We believe that DFT, as well as MLIPs trained on DFT, will continue to have a role in future research at the intersection of ML and materials science but argue that we can utilize MLIPs for more ambitious purposes. Furthermore, recent research works claim and provide reasonable generalization of MLIPs to a set of cases not stricly observed in the DFT-based training data \citep{yang2024mattersim, merchant2023scaling, batatia2023foundation}. These results are encouraging that generalizeable MLIPs can be built, yet the evidence they provide is far from conclusive given conflicting results in other works \citep{bihani2024egraffbench, gonzales2024benchmarking}, indicating the need for further research work. While such challenges may be improved with additional data generation, eventually the limitations of DFT as the source of the data will become the limiation for the underlying MLIP.

\textbf{AV2: Large-Scale Experimental Data is Required to Train ML Models for Materials Discovery.} One of the primary reasons for the success of AlphaFold \citep{jumper2021highly} was the availability of high-quality sequence-structure data obtained by experimental measurements \citep{wwpdb2019protein}. Given the importance of aligning to experimental measurements as the most pertinent ground truth as described in \Cref{sec:mat-limits}, the approach of developing large, ML compatible databases appears attractive. We believe large-scale experimental data collection is a promising idea, yet most of the experimental datasets in materials remain small in large part due to the vast diversity of materials compounds and experimental conditions, as well as the large cost of experimental measurements \citep{xu2023small}. While interesting ideas exist related to driving automated data collection with AI-enabled, self-driving laboratories \citep{miret2024llms, sim2024chemos}, much research remains to be able to scale automated, reproducible experiment execution and data collection in the vast majority of materials design cases. Given this challenge, new research emerged to accelerate the planning, execution and analysis of materials science experiments with machine learning methods \citep{miret2024perspective}. As such, in silico materials design with performant simulation methods will continue to play an important part in materials discovery with MLIPs playing an important role. Furthermore, whereas the initial capabilities of AlphaFold represented a significant breakthrough, further challenges remain in aligning model predictions for real-world conditions \citep{terwilliger2024alphafold}, similar to the ones described in \Cref{sec:materials}. MLIPs integrated with atomistic simulations can be tested against targeted, small-scale experimental measurements, which in turn will drive further understanding of materials behavior.

\textbf{AV3: MLIPs are Not Required for Device Simulations. } Multiscale simulation approaches, which rely mostly on partial differential equations (PDEs) to model different length and time scales, provide a viable alternative to MLIPs for device scale simulation. On top of that, promising approaches for ML-based PDE acceleration are actively being developed \citep{kovachki2023neural, brunton2024promising, brandstetter2022message, takamoto2022pdebench}. We acknowledge that is a valid perspective and that the scalability and accuracy of MLIPs to device scale simulations remains an open research question that requires solving both the engineering and scientific challenges outlined in \Cref{sec:computation}. Evidently, until such challenges can be addressed with new methods, multiscale methods remain the best practical solution. 

\vspace{-0.2cm}
\section{Conclusion}
\vspace{-0.1cm}
For MLIPs to disruptively accelerate materials discovery, the community requires new research priorities that ultimately lead to fully atomistic device scale simulations with quantum mechanical accuracy. These types of simulations will unlock new realms of insight for how the composition of materials affects device performance. Using this guiding principle, we propose a number of interdisciplinary research challenges related to higher quality first-principles data generation, architecture and workflow design, proper inferencing, and algorithmic improvements for computational performance; all of which will require expertise from materials scientists, ML researchers, and software engineers.

%% file: sections/appendix.tex
\appendix
\onecolumn

\section{MLIP Evaluation in Molecular Dynamics Simulation} \label{app:aimd}

\textbf{Ab-Initio Molecular Dynamics (AIMD):} AIMD combines quantum mechanical calculations with classical molecular dynamics to model the dynamics of atoms and molecules. AIMD generally applies a quantum mechanical method, such as DFT or Car-Parrinello Molecular Dynamics (CPMD) \citep{car1985unified}, to compute the electronic energy and atomic forces (the negative gradient of the energy with respect to atom coordinates) which are used to propagate atoms according to classical mechanics. The ability of AIMD toward quantum accurate computation of energy and forces makes it a natural choice for MLIPs deployment shown in a couple of early studies that have indicated limited success \citep{fuforces, bihani2024egraffbench}. AIMD is also a useful method for obtaining correlated training data for MLIPs for real-world applications of materials system under physical conditions, thereby lending itself in enabling device-scale simulations.

\begin{figure}[hbp]
     \vspace{-0.2cm}
    \centering
    \includegraphics[width=0.5\linewidth, trim=0 20 0 20, clip]{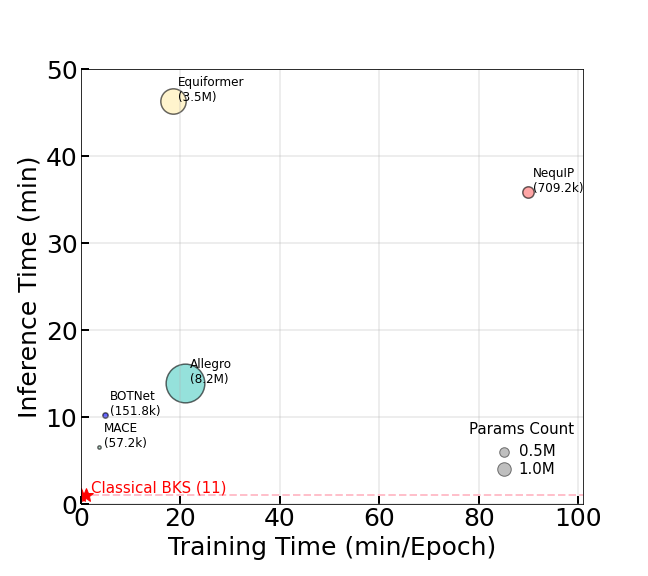}
    \caption{Training and inference times for different MLIP architectures on the LiPS dataset based on an analysis from \citet{bihani2024egraffbench}. The MLIPs architectures include: MACE \citep{batatia2023foundation}, BotNet \citep{batatia2025design}, Allegro \citep{musaelian2023learning}, Equiformer \citep{liao2023equiformer} and NequiP \citep{batzner20223}, all of which fail to achieve the inference time performance of the classical BKS potential \citep{van1990force}.}
    \label{fig:model_scaling}
\end{figure}

\textbf{MLIP Inference and Training Time Comparison:} \Cref{fig:model_scaling} based on \citet{bihani2024egraffbench} illustrates the inference and training times for several geometric deep learning architectures developed for molecular dynamics (MD) simulations. It is important to note that while these models often have a large number of parameters, leading to higher inference times, certain architectures—such as transformers—are relatively fast during inference. However, transformer-based models require more epochs for training, which increases the overall training cost.

\section{MLIP Evaluation at Large Scale for Diverse Materials} \label{app:sec:mlip-eval}

\begin{figure}
\centering     
\subfigure[Normalized Element Frequency in Log Scale]{\label{fig:a}\includegraphics[width=0.45\textwidth]{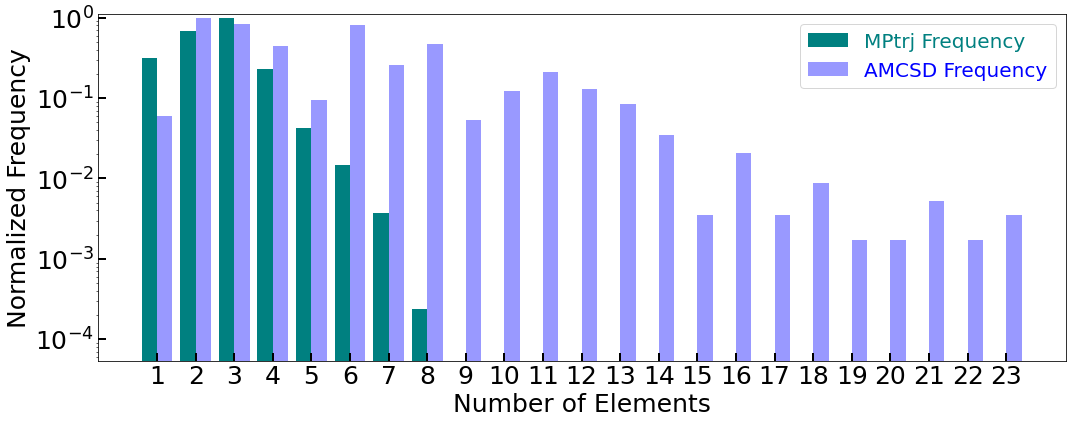}}
\subfigure[Normalized Element Frequency in Linear Scale]{\label{fig:b}\includegraphics[width=0.45\textwidth]{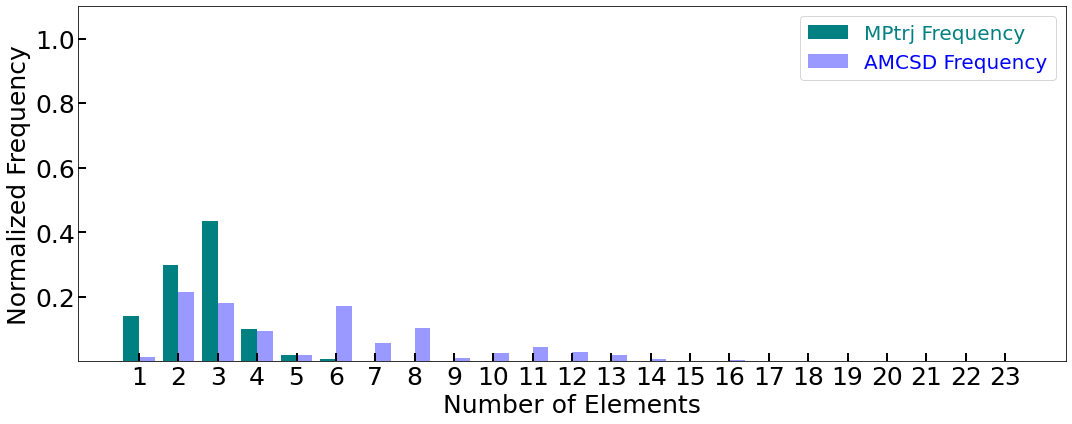}}
 \vspace{-0.40cm}
\caption{Normalized element frequency of Mptrj \citep{deng2023chgnet} containing DFT simulated and naturally occuring minerals in AMCSD \citep{downs2003american} that are experimentally measured compounds. The linear scale shows the high proportion of Mptrj related to binary and ternary materials. }
\label{fig:amcsd_main}
\end{figure}

Our analysis in \Cref{sec:mat-limits} shows the limited space of coverage of common DFT datasets, such as Mptrj \citep{deng2023chgnet}. The shortage of complex materials can further be augmented by comparing Mptrj to experimental databases, such as real-world experimentally measured minerals from the AMCSD database \citep{downs2003american}. One observation is that real minerals consist of far more components than just binary and ternary crystals as shown in \Cref{fig:amcsd_main}. While the maximum number of elements in Mptrj is 9, AMCSD contains meaningful representation with much larger number of elements, with some containing up to 23 components. This highlights that the interactions between such complex compositions might not be captured by the universal potential trained only on the MPtrj dataset, suggesting that these materials may be considered out of distribution.

\subsection{Considerations for Large-Scale MD Bechmarking}

The utility of MLIPs span a broad application space, from single system inference, to large scale material screening. In the way that computer vision models are just as easily applied to one image as they are to a collection of real time video feeds, and single point weather forecasting is transferable to global scale weather modeling, we would like to be able to apply MLIPs across this range of system scale while maintaining model consistency in with respect to both scientific and engineering considerations. To get to this level of robustness for MLIPs requires many interrelated considerations, starting from base level materials and working towards device scale simulations. In base level materials, the specific system will inform important simulation parameters that underlie the behavior of the model. These simulation parameteres include boundary conditions, data preprocessing pipelines, convergence thresholds, as well as downstream evaluation metrics, all of which are conditioned on the base material. Many of these scientific parameters have concrete computational impacts, and when evaluating systems at scale, can have major implications for simulation throughput. While simulation parameters and compute requirements can be meticulously tuned when evaluating individual systems, fine-tuning quickly becomes intractable with large-scale evaluation of diverse materials. 

Furthermore, different MLIPs may have different computational costs, creating further complexity when comparing MLIPs at large scale. Practically, engineering considerations and compute constraints need to be considered and reported when developing broad MLIP evaluation pipelines. Similar to consistency issues with DFT described in \Cref{sec:dft-limits}, further work is also needed in providing consistency, transparency and reproducibility in MLIP evaluation. For example, most models come with their own \textit{calculator} interface designed to plug into virtual material modeling environments which orchestrate simulations. These calculator implementations are not readily transferable between models, and may include assumptions and nuanced model and data specific transforms, rendering one model's calculator incompatible with other models. This implementation detail makes direct simulation comparisons difficult unless the explicit calculator paradigms are explicitly provided and comparable in their implementation. 

Additionally, MLIP data processing pipelines can have subtle differences that can make reproducing results challenging. Some models may expect fractional coordinates, while others rely on Cartesian, and in the case of GNN's, the graph creation process itself may vary amongst data processing pipelines. When further extended to a framework to framework comparison, common graph libraries such as PyG \citep{Fey/Lenssen/2019} and DGL \citep{wang2019dgl}, and two of the most popular material analysis libraries, pymatgen \citep{ong2013python}, ASE \citep{HjorthLarsen_2017} and LAMMPS \citep{thompson2022lammps}, may lead to divergent results simply due to convention of implemented functions. As such, to evaluate a collection of MLIPs consistently, common frameworks, calculator interfaces, computational constraints, and representation formats should all be considered. Current benchmarks, such as Matbench \citep{riebesell2023matbench}, fall short in properly evaluating the performance of MLIPs in these simulation settings. For example, while many models show similar performance in test metrics such as energy and force MAE, they show great variance when used in basic molecular dynamics tasks such as crystal structure relaxation \citep{gonzales2024benchmarking, bihani2024egraffbench}, showing that strong static benchmarking results do not always correlate with simulation performance. Given the importance evaluation to ML method development, integrating MLIPs into simulations remains critical to enable greater understanding of the underlying materials.

\section{Model \& Data Distillation Methods} \label{app:distillation}

\textbf{Data distillation:} Data distillation in MLIPs represents a critical, yet underaddressed, challenge, wherein strategic configuration selection from large-scale datasets (such as MPTrj, OMat24, Alexandria) can substantially reduce computational training costs. Existing active learning approaches insufficiently address the complex multi-dimensional challenge of identifying representative, information-rich configurations that capture the configurational space's essential physics. Moreover, in active learning approaches such as the full retraining approach in \citep{gonzales2023data}, the computational cost is fundamentally driven by the requirement of retraining the entire model in each iteration, which becomes prohibitively expensive when comprehensive datasets are available a priori. Systematic sampling methodologies must balance configuration diversity, statistical representativeness, and information entropy to enable efficient potential training. The goal is to develop principled strategies that can extract minimal yet maximally informative subsets from massive trajectory databases, potentially reducing computational cost of training by orders of magnitude.

\textbf{Model distillation:} Model distillation is an emerging strategy for scaling MLIPs to device-level simulations through knowledge transfer from large ``teacher'' models to compact ``student'' models exploiting, for instance, energy hessians~\cite{amin2025towards}. This approach enables the development of parameterically efficient models that can potentially surpass their large-scale predecessors in system-specific performance. The core methodology involves systematically compressing the knowledge representation of complex MLIPs into smaller neural network architectures. These distilled models can be further refined on targeted, minimal datasets, leveraging their reduced parameter space to achieve computational efficiency comparable to classical potentials while maintaining high-fidelity predictive capabilities. However, further investigation is required on the distilling strategy, the stability and performance of the consequent MLIP, including its generalizability. 